\documentclass[preprint2]{aastex}

\shorttitle{}
\shortauthors{}

\begin{document}


\title{On the nature of Off-pulse emission from pulsars}


\author{Rahul Basu$^1$, Dipanjan Mitra$^1$ and Ramana Athreya$^2$}
\affil{$^1$National Centre for Radio Astrophysics, P. O. Bag 3, Pune University
Campus, Pune: 411 007. India.\\
$^2$Indian Institute of Science Education \& Research (IISER) - Pune, 900, NCL Innovation Park, Homi Bhabha Road, Pune: 411 008. India.}
\email{rbasu@ncra.tifr.res.in, dmitra@ncra.tifr.res.in, 
       rathreya@iiserpune.ac.in}


\begin{abstract}
\noindent
In Basu et al. 2011 we reported the detection of Off-pulse emission from 
two long period pulsars B0525+21 and B2045--16. The pulsars were observed 
at a single epoch using the 325 MHz frequency band of the Giant Meterwave 
Radio Telescope (GMRT). In this paper we report a detailed study of the 
Off-pulse emission from these two pulsars using multiple observations at 
two different frequencies, 325 MHz and 610 MHz bands of GMRT. We report 
detection of Off-pulse emission during each observation and based on the 
scintillation effects and spectral index of Off-pulse emission we conclude 
a magnetospheric origin. The magnetospheric origin of Off-pulse emission 
gives rise to various interesting possibilities about its emission mechanism 
and raises questions about the structure of the magnetosphere.

\end{abstract}





\section{\large Introduction}

The basic pulsar phenomenon where a series of highly regular narrow pulses is
observed has been explained from the very outset by invoking the lighthouse 
model. In this model the highly magnetized, rotating neutron star has a charge 
filled magnetosphere. The magnetic field surrounding the central star is 
characterized by a dipolar field with open and closed field line regions. 
The locus of the foot prints of the open field lines on the neutron star 
surface constitute the polar cap. It is believed that plasma is accelerated 
to relativistic velocities at the polar cap, which streams out along the 
open field lines giving rise to the observed coherent radio emission. This 
originate a few hundred kilometers above the stellar surface, and appear as 
narrow pulses due to the rotation of the neutron star. The plasma remains 
trapped in the closed field line regions thereby abstaining from the emission 
process.

The known cases of emission from pulsars outside the main pulse are the 
interpulse, pre/post-cursor emission and the pulsar wind nebulae (PWNe). The 
interpulse and pre/post-cursors emission appear as temporal structures in the 
time series data. The interpulse emission (located $180\degr$ away from main 
pulse) are usually associated with orthogonal rotators, the radio emission 
in the interpulse believed to originate from the opposite magnetic pole. The 
pre/post-cursor emission are located closer to the main pulse and connected 
via a bridge emission. They are highly polarized and their location in the 
magnetosphere is uncertain. The time series observations of pulsars are 
insensitive to any temporally constant background emission. The technique of 
gated interferometry, where the main pulse emission is blocked, can probe 
a constant background emission by mapping the Off-pulse region. In the past 
gated interferometric studies have been employed to detect PWNe from very 
energetic young pulsars with spindown energies $\dot{E} \geq$ $10^{34}$ erg 
s$^{-1}$~\citep{fra97} .

In recent studies we reported the detection of off-pulse emission from 
two long period pulsars based on gated interferometric observations with 
the Giant Metrewave Radio Telescope (GMRT) at 325 MHz~\citep{bas11},
(hereafter PaperI). Our detection of off-pulse emission is a `unique' 
and surprising result because the two long period pulsars are relatively 
old and less energetic unlike the pulsars known to harbour PWN. Also the 
pulse profiles do not show any temporal structures outside the main pulse. 
We concluded on heuristic grounds that the detected Off-pulse emission is 
magnetospheric in origin based on outlandish expectations of ISM densities 
to sustain a PWM for these pulsar energetics.  
 
In this paper we continue our previous investigations of Off-pulse emission. 
In section~\ref{section2} we report the detection of Off-pulse emission at 
widely separated times and frequencies (see section~\ref{subsec2.1}) and 
also account for the authenticity of these detections by carrying out 
a detailed test of the instrument to weed out possibility of the detected 
emission being a result of smearing of the pulsed signal across time (see 
section~\ref{subsec2.2}). In section~\ref{section3} we study the flux 
variations which lead us to an upper limit of the size of the Off-pulse 
emission region (see section~\ref{subsec3.1}) and we also obtain secure 
estimates of its spectral index $\alpha$ (see section \ref{subsec3.2}). 
Based on our estimates of these properties we conclude a magnetopsheric 
origin for Off-pulse emission. Finally we discuss the results and their 
implications in section~\ref{section4}.

\begin{table*}
\begin{center}
\caption{ Observation details
\label{Obs_detail}}
\footnotesize{\begin{tabular}{cccccccccc}
\tableline\tableline
Date & Pulsar & freq(chan) & Time & time res & Corr &         ang res       & On flux(rms) & Off flux(rms) &$\frac{Off}{On}$ \\
     &        &    MHz     & mins &   msec   &      & \arcsec$\times$\arcsec&     mJy      &      mJy      &                 \\
\tableline
18 Jan, 2010*& B0525+21& 325(128) & 160 & 261 &GHB& 9.5$\times$6.5& 30.0$\pm$2.1(0.55)&3.9$\pm$0.5(0.45)&0.130$\pm$0.026\\
08 Jul, 2011 & B0525+21& 325( 32) & 240 & 250 &GSB& 8.3$\times$6.7& 44.5$\pm$2.4(0.65)&6.6$\pm$0.5(0.4)&0.148$\pm$0.019\\
09 Jul, 2011 & B0525+21& 325( 32) & 120 & 250 &GSB&10.0$\times$7.4& 48.2$\pm$2.6(0.7)&6.5$\pm$0.6(0.5)&0.135$\pm$0.020\\
15 Feb, 2011 & B0525+21& 610( 32) & 330 & 250 &GSB& 4.8$\times$4.0& 21.1$\pm$1.5(0.45)&3.6$\pm$0.3(0.13)&0.171$\pm$0.026\\
22 Jul, 2011 & B0525+21& 610( 32) & 240 & 250 &GSB& 5.4$\times$3.7& 20.0$\pm$1.2(0.65)&2.8$\pm$0.2(0.13)&0.140$\pm$0.018\\
             &               &     &     &   &               &             &           &               \\
16 Jan, 2010*&B2045--16& 325(128) & 180 & 131 &GHB&11.9$\times$7.2&110.5$\pm$7.9(2.1)&4.3$\pm$1.1(0.65)&0.039$\pm$0.013\\
03 Aug, 2011 &B2045--16& 325(256) & 200 & 125 &GSB&10.2$\times$7.7& 57.5$\pm$3.2(0.35)&6.8$\pm$0.4(0.2)&0.118$\pm$0.014\\
14 Feb, 2011 &B2045--16& 610(128) & 180 & 131 &GHB& 5.8$\times$4.3& 23.9$\pm$1.7(1.0)&1.1$\pm$0.2(0.13)&0.046$\pm$0.012\\
25 Aug, 2011 &B2045--16& 610(256) & 140 & 125 &GSB& 4.9$\times$4.3& 43.2$\pm$3.1(0.45)&4.6$\pm$0.3(0.11)&0.106$\pm$0.015\\

\tableline\tableline
\end{tabular}}
%% Any table notes must follow the \end{tabular} command.
\tablenotetext{~}{\small * reported in Basu et al. 2011}
\end{center}
\end{table*}

\begin{figure}
\includegraphics[angle=0,scale=.5]{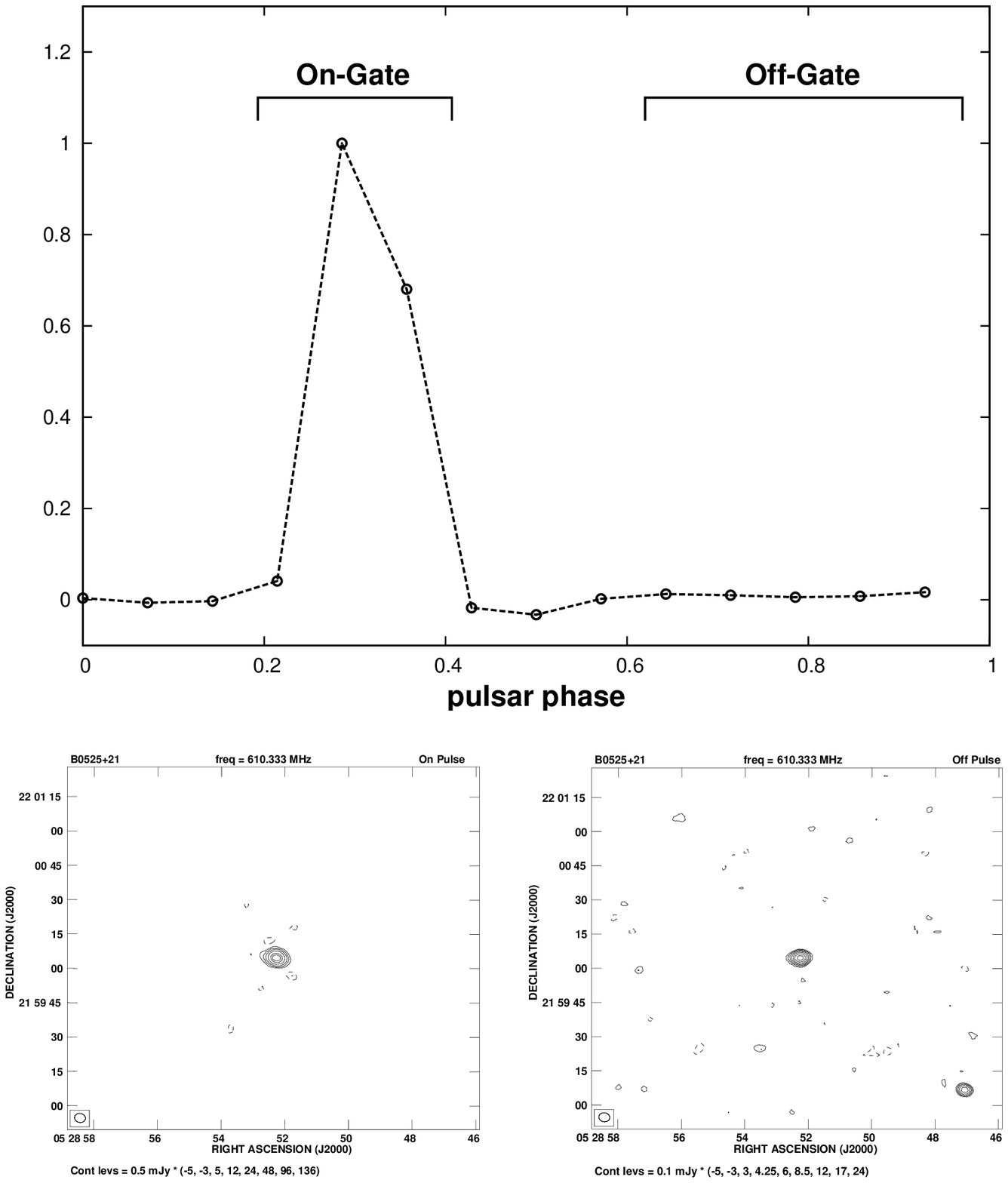}
\caption{The folded profile (top) of B0525+21 representing the On- and
Off-pulse gates. The corresponding contour maps of the On-pulse (bottom left)
and Off-pulse (bottom right) emission from the pulsar. The data was recorded
on 15 February, 2011 at 610 MHz.
\label{fig_P1imag}}
\end{figure}

\begin{figure}
\includegraphics[angle=0,scale=.5]{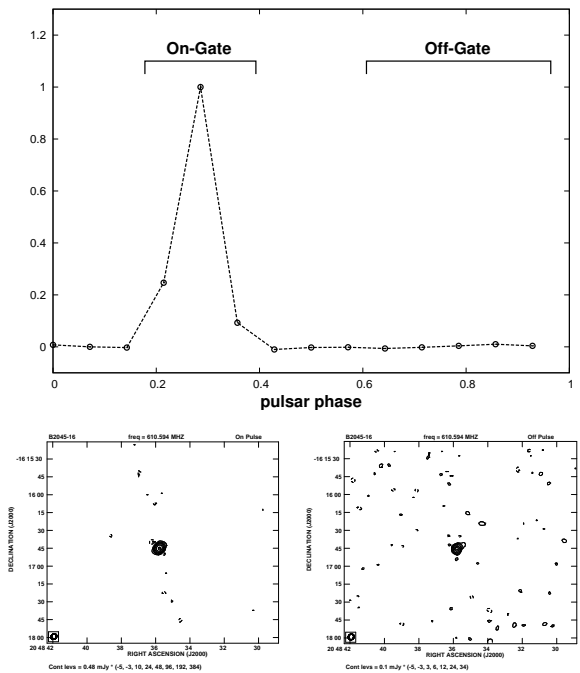}
\caption{The folded profile (top) of B2045-16 representing the On- and
Off-pulse gates. The corresponding contour maps of the On-pulse (bottom left)
and Off-pulse (bottom right) emission from the pulsar. The data was recorded
on 25 August, 2011 at 610 MHz.
\label{fig_P2imag}}
\end{figure}

\section{\large Observations \& Analysis}
\label{section2}

\subsection{Multi-epoch/ Multi-frequency observations}
\label{subsec2.1}
We reported detection of off-pulse emission in paperI from two pulsars PSR 
B0525+21 and B2045--16 based on single epoch observations at 325 MHz with 
the GMRT. We have now observed these two pulsars at 325 MHz and 610 MHz 
multiple times. The observations were carried out between January 2010 and 
August, 2011 (see table~\ref{Obs_detail}). The specialized interferometric 
high time resolutions of 128/256 milliseconds supported by the GMRT correlator 
system were used for these observations. We observed the pulsars at both 325 
MHz and 610 MHz frequency band. During the early months of 2011 the GMRT 
correlator system was upgraded from the previously existing Hardware backend 
(GHB) to a Software backend (GSB) which is currently under use. The GHB was 
used with the lowest available time resolution of 131 msec and a bandwidth 
of 16 MHz split into 128 channels. The GSB initially operated at a lowest 
time resolution of 251 milliseconds over a 16 MHz bandwidth spread across 
32 channels. The GSB was extended to operate at 125 milliseconds time 
resolution over 16 MHz bandwidth split into 256 channels. All these modes 
were used for our observations as documented in columns 1 to 6 of 
table~\ref{Obs_detail}.

We used the technique of `offline-gating' discussed in paperI to image the 
On- and Off-pulse regions. To summarize briefly, the interferometric time 
series data was folded with the periodicity of the pulsar to determine the 
On- and Off-pulse phases; a phase was assigned to each time record and two 
`gates' were put in to separate the On- and Off-phase data, which were then 
imaged and the flux of the point source coincident with the pulsar position 
in each image measured. To check for the consistency of measured pulsar flux 
over different epochs of observations we compared the flux of the other point 
sources in the field which were within 10\% (within calibration errors) for 
all observations. This indicates that any large variations in the On- and 
Off-pulse flux were intrinsic effects.

Table~\ref{Obs_detail} gives the details of the experimental setup and also 
summarizes the results of the analysis. Column 7 shows the angular resolution 
in the images. During each of these observations a point source was detected 
at the location of the pulsars in both the On- and Off-pulse maps. The On-pulse
flux, averaged over the pulse period (all On-pulse flux reported in this paper
are averaged over the entire period), are reported in column 8 while column 9
report the measured Off-pulse flux. The ratio of the Off-pulse flux and 
period averaged On-pulse flux is shown in column 10. An example of Off-pulse 
and On-pulse image for a single epoch at 610 MHz is shown in 
figure~\ref{fig_P1imag} and~\ref{fig_P2imag}.

Thus in summary, in each of the multi-epoch and multi-frequency interferometric 
observations we have detected off-pulse emission from PSR B0525+21 and 
B2045--16. In our studies we used two different correlator systems, the GHB 
and GSB, and two different frequency bands, 325 MHz and 610 MHz bands, which 
make it extremely unlikely for the Off-pulse to be a spurious detection.

%% If you use the table environment, please indicate horizontal rules using
%% Do not put multiple tabular environments within a single table.
%% The optional \label should appear inside the \caption command.

%% In this section, we use  the \subsection command to set off
%% a subsection.  \footnote is used to insert a footnote to the text.


%% This section also includes several of the displayed math environments
%% mentioned in the Author Guide.

\subsection{\large Pulsed Noise Source: Smearing of signal across time.}
\label{subsec2.2}
In paperI, we discussed the possibility that off-pulse emission was a
result of temporal leakage of the pulsed signal. To investigate this 
we recorded front end terminated 131 milliseconds interferometric data 
and estimated the autocorrelation of the noise signal for each baseline. 
We found that the observed autocorrelations were significantly lower (0.04\%) 
than the required levels (around 1\%) to explain the observed off-pulse 
emission (see paperI for details).

However, the above experiment did not account for temporal leakages due to a 
strong pulsed signal as seen in pulsars. In order to investigate this effect 
a series of narrow single pulses with no Off-pulse emission was required to 
be introduced in the receiver system. A broad-band pulsed noise source with 
a period of 4 seconds and a duty cycle of 32 millisecond was developed for 
this purpose. The power of the On-pulse signal was -5 dbm and the signal was 
down by 30 db in the Off-state, which was significantly below the detected 
Off-pulse emission. The noise source was radiated from an antenna in the 
central square of the GMRT. Interferometric data was recorded with a time 
resolution of 251 milliseconds for about half an hour with the noise source 
switched on. Since the noise source was in the near field, the geometrical 
delay calculations were vastly complicated (it is a function of the distance 
of the individual antennas from the pulsed noise source). We did not correct 
for any geometrical delay in the correlator chain. This implied that only for 
a handful of cases (10 baselines) the antennas were close enough for the 
coherence condition to hold and these baselines were used for the subsequent 
analysis as discussed below.

If there was no leakage of the pulsed signal in time, the folded profile with
increasing number of periods averaged together, for any baseline, would show
a decrease in the off-pulse noise and an increase in the signal to noise ratio
(SNR) by a factor $\sqrt{N}$, where N is the number of folded periods. 

The interferometer measures the correlation between the voltages recorded by 
the individual elements~\citep{tho86}.\\\\
$ r(\tau) = \frac{1}{2T}\int_{-T}^{T} V_{1}(t)\times V_{2}(t-\tau)~dt$\\\\
where the quantity $r(\tau)$ is related to the intensity of the incident 
signal and 2T is the temporal resolution (251 msec). 

We assume that the system introduces a constant fractional leakage $\epsilon$ 
into the adjacent time bin (this is the most likely situation, besides any 
other situation will lead to a lower SNR). The correlation in the adjacent 
bin in such a scenario is given as:\\\\
$  r'(\tau) = \epsilon^{2}\frac{1}{2T}\int_{-T}^{T} V_{1}(t)\times V_{2}(t-\tau)~dt, ~~\epsilon < 1 $ \\\\
The statistics of the bins adjacent to the pulsed source is characterized by 
a mean($\mu$) and rms ($\sigma$) given as:
\begin{equation}
\mu = \epsilon^{2n}~\frac{1-\epsilon^{2N}}{N~(1-\epsilon^{2})}\times r(\tau)
\label{eqn_corr1}
\end{equation}
\begin{equation}
\sigma \sim \frac{\epsilon^{2n}}{(1-\epsilon^{2})}~\sqrt{\frac{1}{N}\frac{1-\epsilon^{2N}}{1+\epsilon^{2}}}\times r(\tau)
\label{eqn_corr2}
\end{equation}
Here the first bin is the $n^{th}$ bin from the pulsed source with N bins 
used for statistics; $r(\tau)$ is the correlation in the pulsed source bin. 
The statistics of the adjacent bins will be governed by a combination of 
gaussian statistics due to random noise and the constant leakage into adjacent 
bins. However for sufficient long averaging the effect of random noise will 
be overshadowed by the leakage term and the SNR for the pulsed source would 
saturate at a constant level.

\begin{figure*}
\begin{tabular}{@{}lr@{}}
{\mbox{\includegraphics[height=5cm,width=8cm,angle=0.]{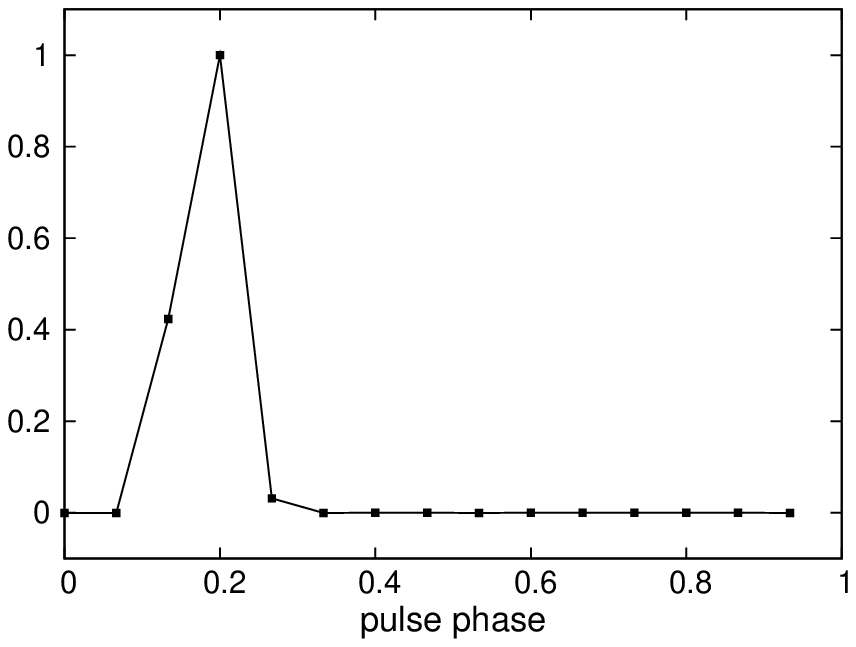}}} &
{\mbox{\includegraphics[height=5cm,width=8cm,angle=0.]{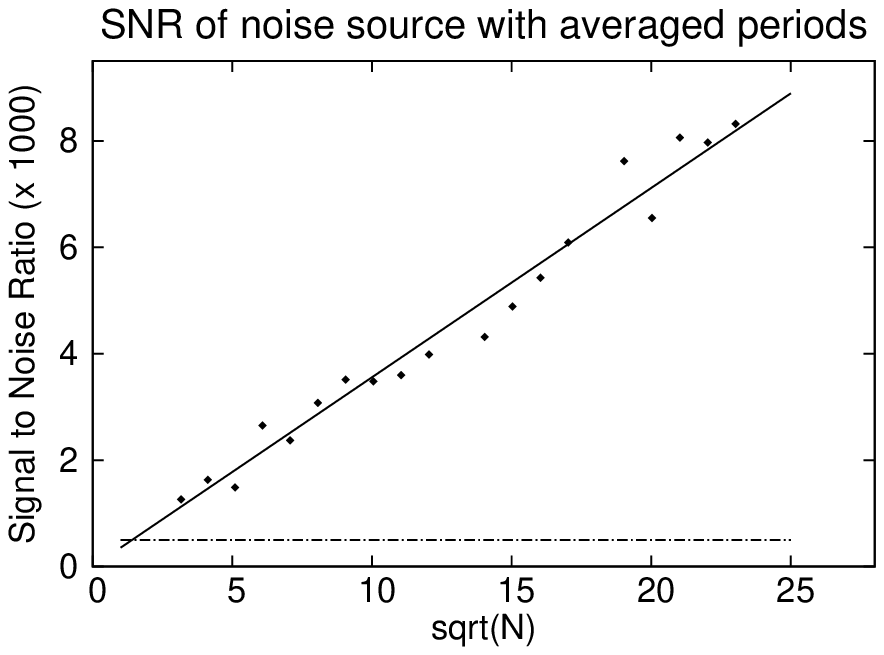}}} \\
\end{tabular}
\caption{\small The pulsed noise source as observed with the interferometer
(left), with 565 folded periods of 4 sec and a time resolution of 250 msec.
The change in signal to noise ratio of the pulsed signal with increasing 
number of folded periods (N) from 10 periods to 565 periods (right). The dark 
solid lines represent a linear increase of SNR with $\sqrt{N}$ indicating the
baseline to be noise like. The dot dashed horizontal line is the level at 
which the SNR was expected to saturate if temporal leakage was the source 
of Off-pulse emission.}
\label{fig_Nfold}
\end{figure*}

The interferometric data from the pulsed noise source were folded with a
periodicity of 4 seconds to determine the profile (see fig~\ref{fig_Nfold}). 
A large number of profiles of the noise source were generated by folding an 
increasing number of periods. The SNR for each of these profiles were 
calculated, by dividing the total signal in the noise bin with the rms 
fluctuation in the adjacent bins. In figure~\ref{fig_Nfold} we plot the SNR 
as a function of the square root of the number of periods (points) along 
with the best fit (dark solid line).

We have detected off-pulse emission in pulsars at a level of 0.5 -- 1 \% of 
the On-pulse flux. The Off-pulse region in our studies are 5 bins (n=5) from 
the On-pulse and we can calculate the leakage ($\epsilon$) required using 
equation~\ref{eqn_corr1} and \ref{eqn_corr2} to explain the detected off-pulse.
This would lead to a SNR $\sim$ 500, as represented by the dot dashed 
horizontal line in figure~\ref{fig_Nfold}. However the bins adjacent to the 
noise source continue to follow gaussian statistics at SNR $\sim$ 8000 as 
seen in figure~\ref{fig_Nfold}. This implies that the temporal leakage can 
only lead to a Off-pulse emission which is less than 0.03\% of the On-pulse 
flux. Hence the detected off-pulse emission cannot originate as a result of 
temporal leakage of the On-pulse signal into adjacent time bins.

The unprecedented detection of Off-pulse emission makes it imperative to 
exhaust all possibilities of spurious detection. In this section we report 
repeated detections of off-pulse emission in multiple epochs and frequencies 
using different correlator systems and also rule out spurious detection that 
can result due to temporal leakage. We have established that the Off-pulse 
emission is genuine and not a result of systematic anomaly. We now look ahead 
to determining the nature of Off-pulse emission.

\section{\large Magnetospheric origin of off-pulse: Flux Variation \& Spectral 
Index study.}
\label{section3}
\begin{figure*}
\begin{tabular}{@{}lr@{}}
{\mbox{\includegraphics[height=5cm,width=8cm,angle=0.]{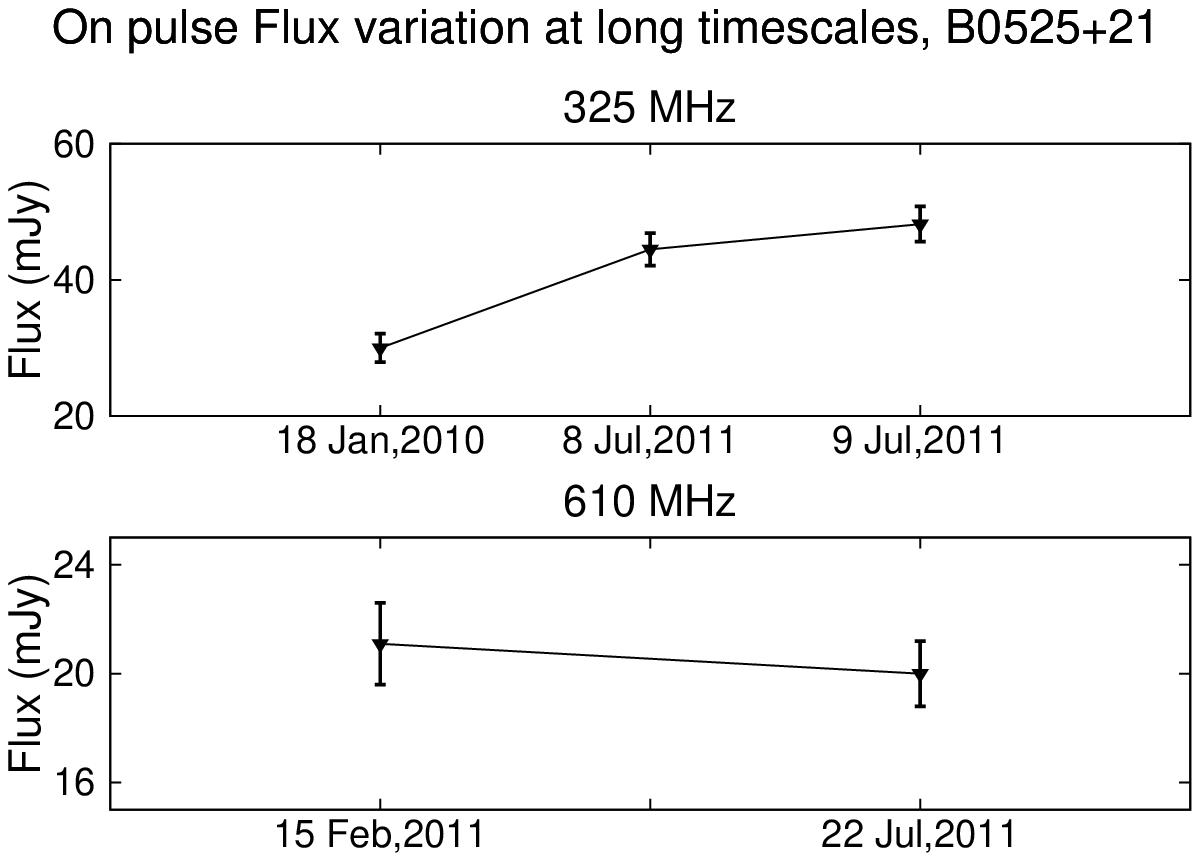}}} &
{\mbox{\includegraphics[height=5cm,width=8cm,angle=0.]{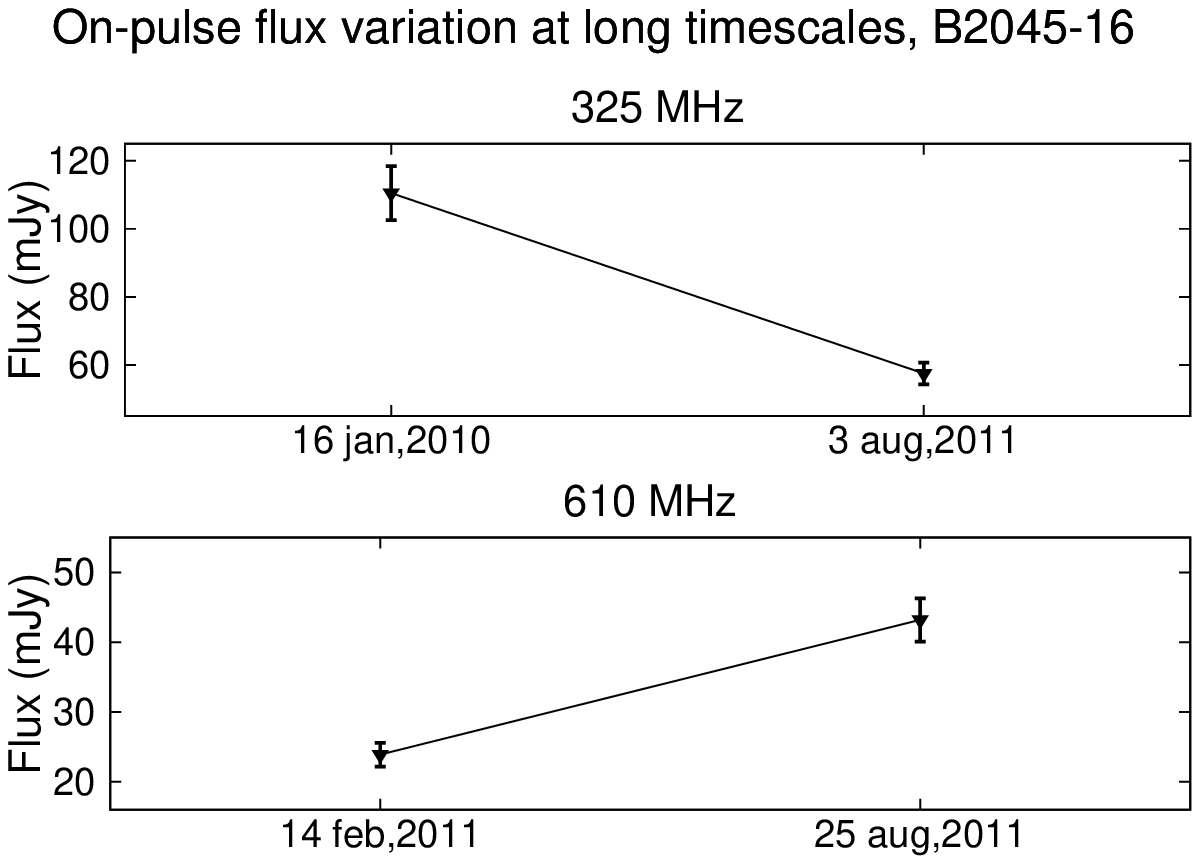}}} \\
{\mbox{\includegraphics[height=5cm,width=8cm,angle=0.]{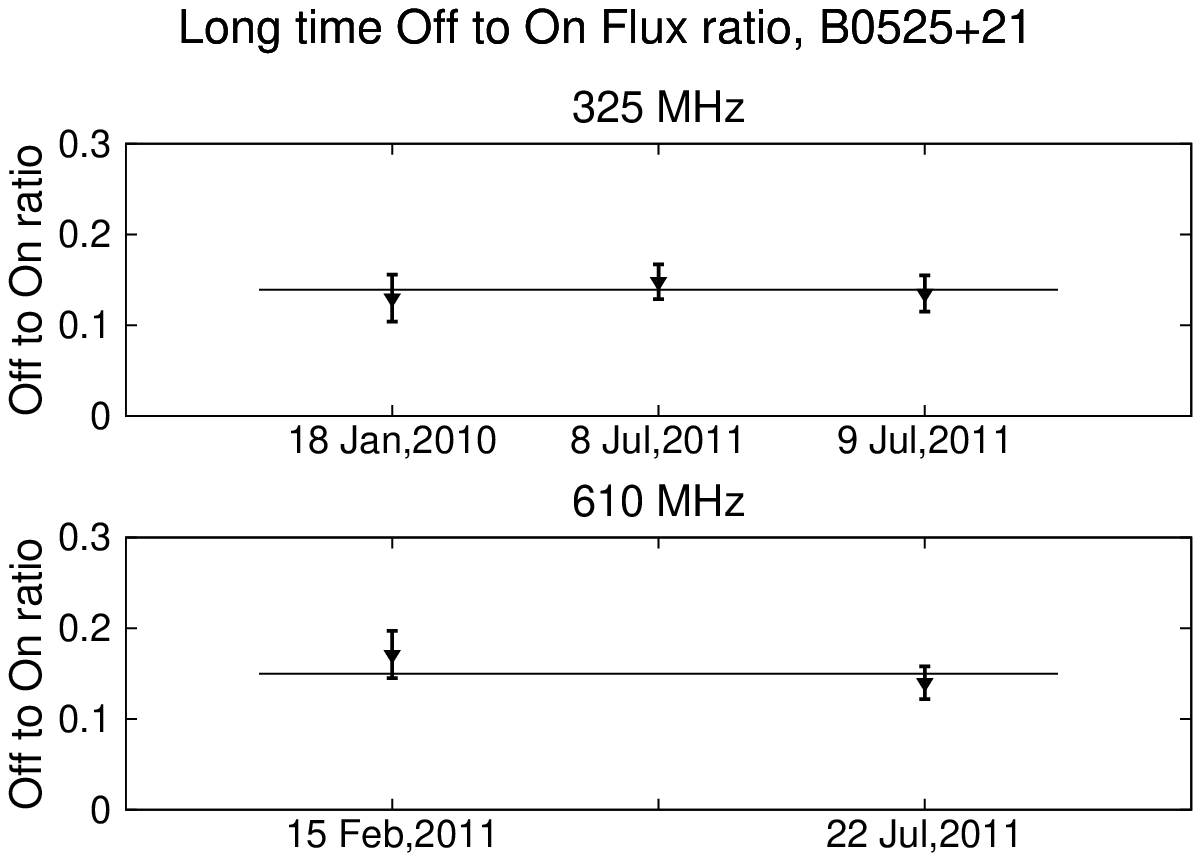}}}&
{\mbox{\includegraphics[height=5cm,width=8cm,angle=0.]{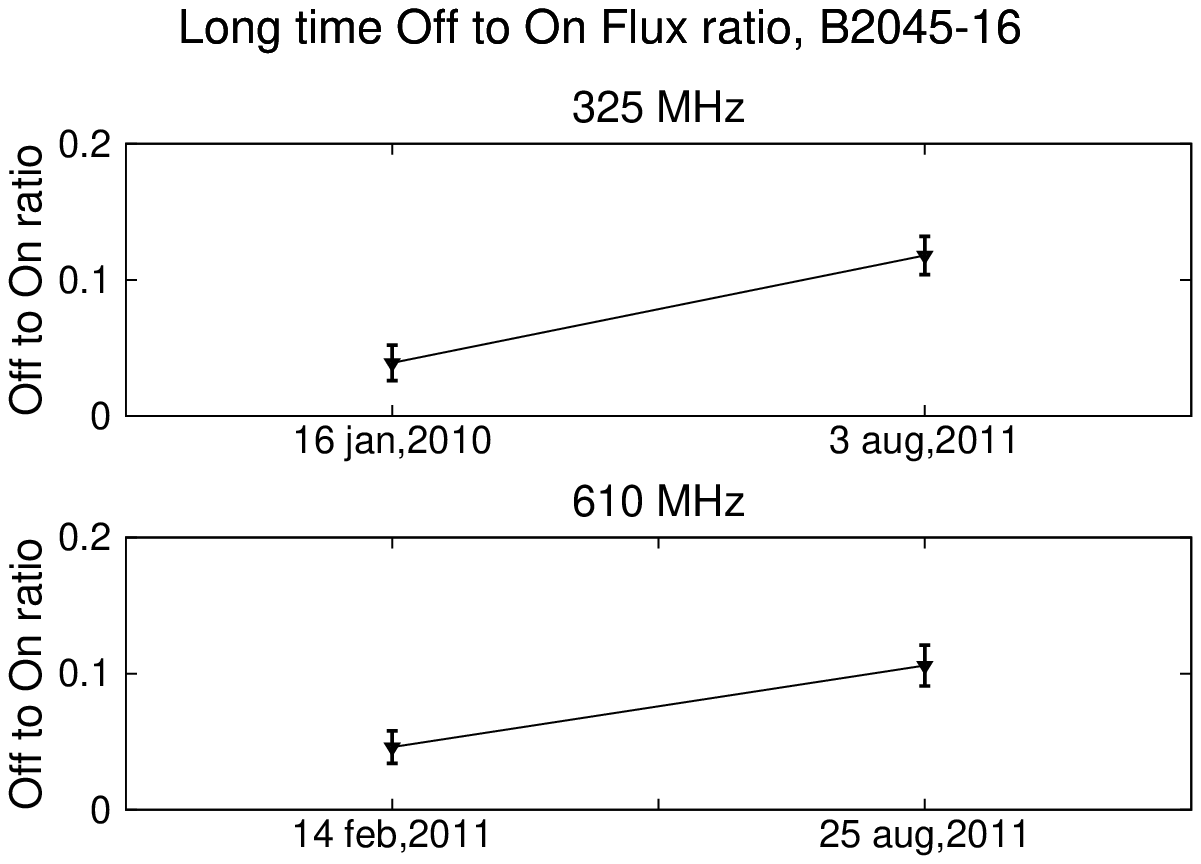}}}\\
\end{tabular}
\caption{\small The plots show variation of the On-pulse flux and the Off-pulse 
to On-pulse flux ratio at large timescales. Each point in the plots is obtained
averaging a single observing run lasting a few hours with the observations 
being distributed over several months. The pulsar B0525+21(left) show a 
constancy in the Off-pulse to On pulse flux ratio over these timescales, while
for B2045--16(right) we observe an apparent variation.}
\label{fig_long}
\end{figure*}

\begin{figure*}
\begin{tabular}{@{}lr@{}}
{\mbox{\includegraphics[height=4.5cm,width=7.2cm,angle=0.]{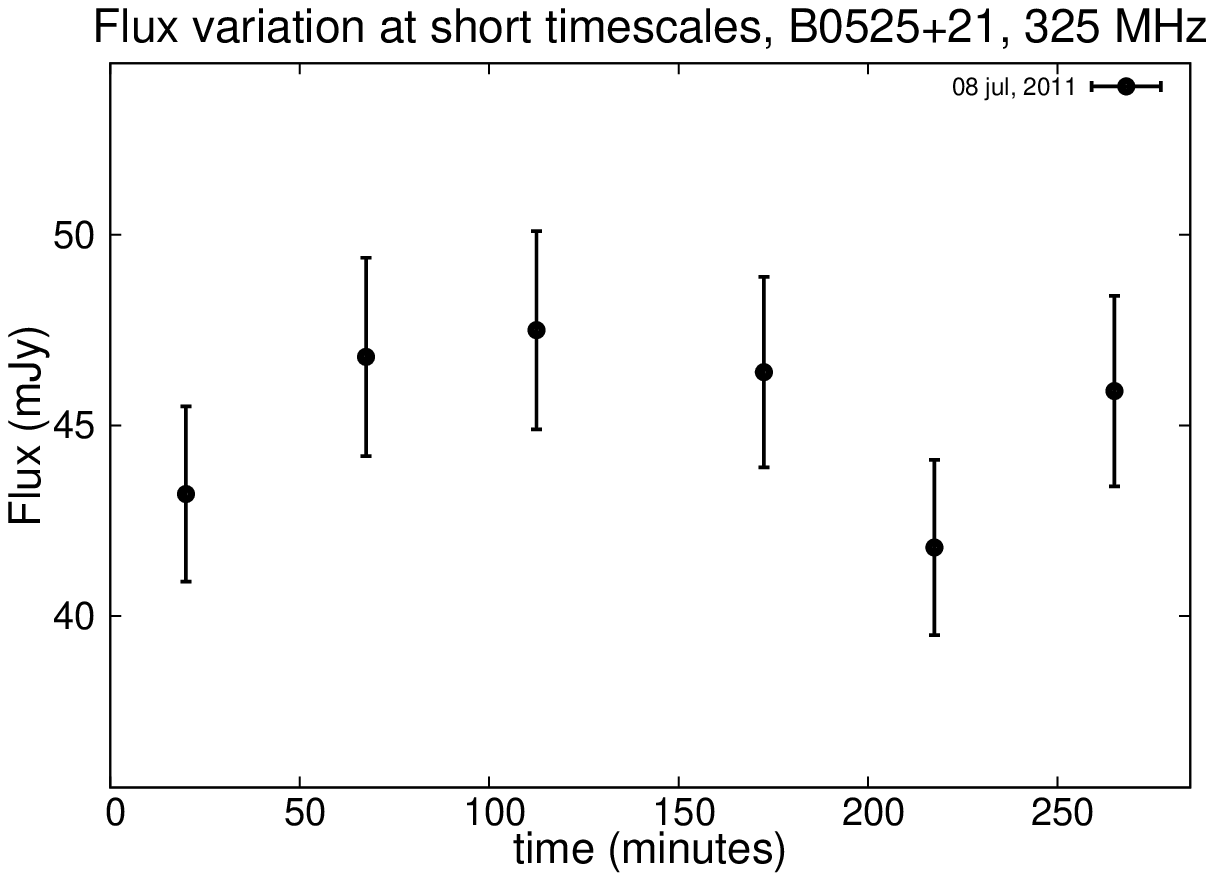}}}&
{\mbox{\includegraphics[height=4.5cm,width=7.2cm,angle=0.]{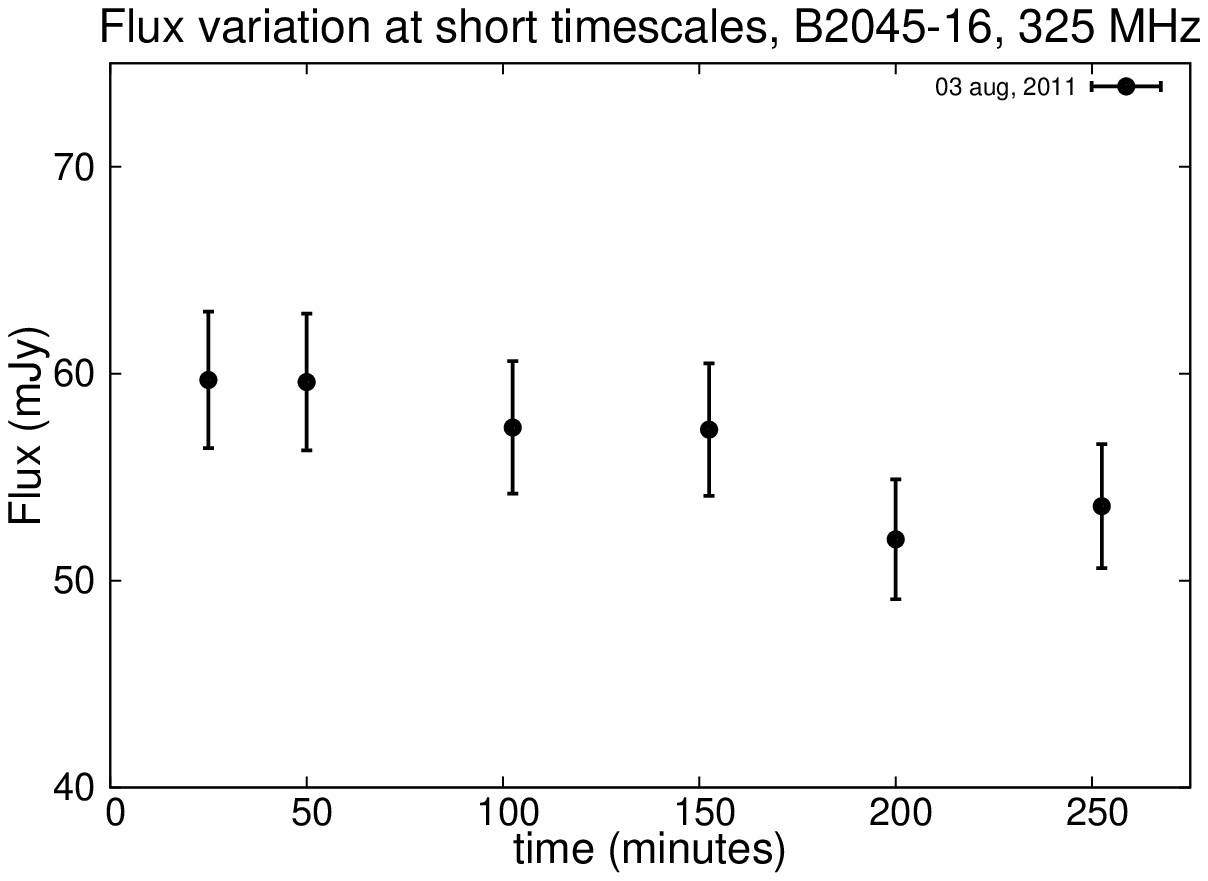}}}\\
{\mbox{\includegraphics[height=4.5cm,width=7.2cm,angle=0.]{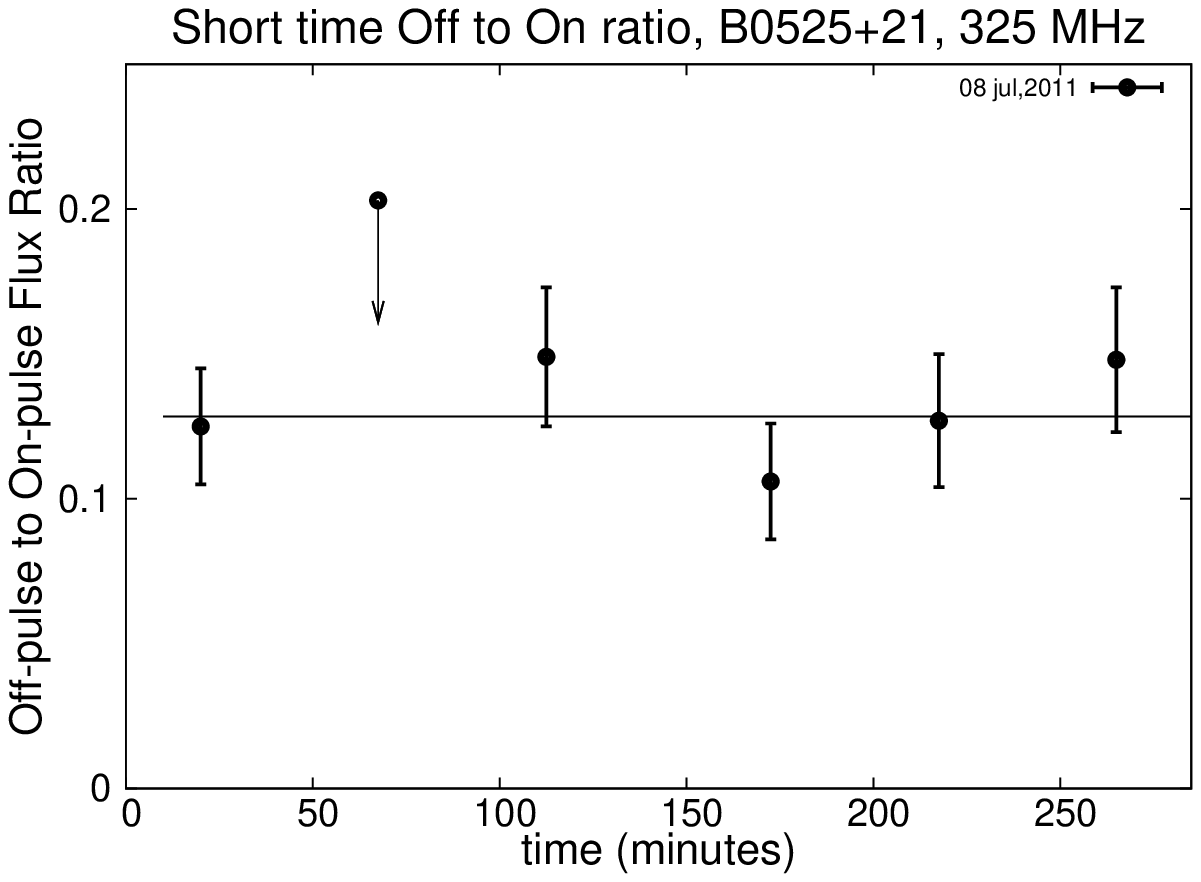}}}&
{\mbox{\includegraphics[height=4.5cm,width=7.2cm,angle=0.]{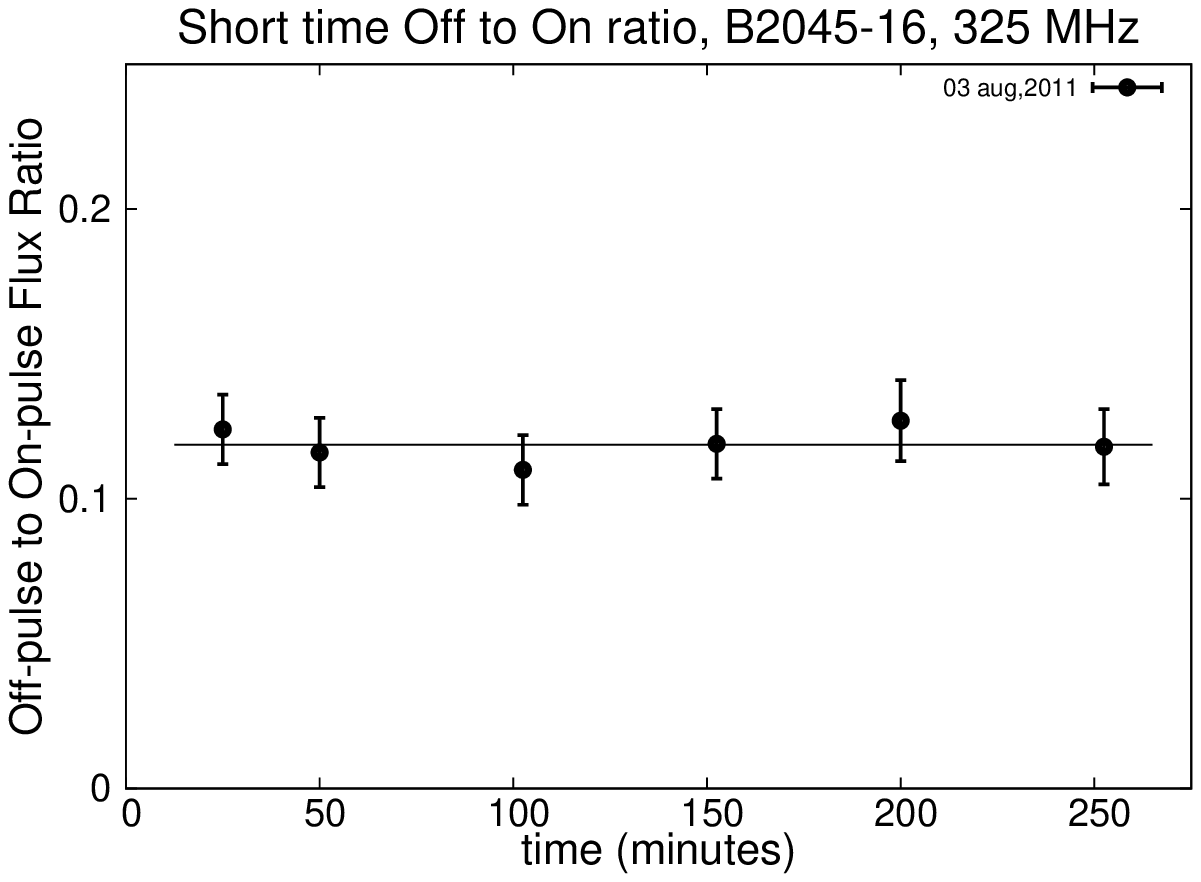}}}\\
{\mbox{\includegraphics[height=4.5cm,width=7.2cm,angle=0.]{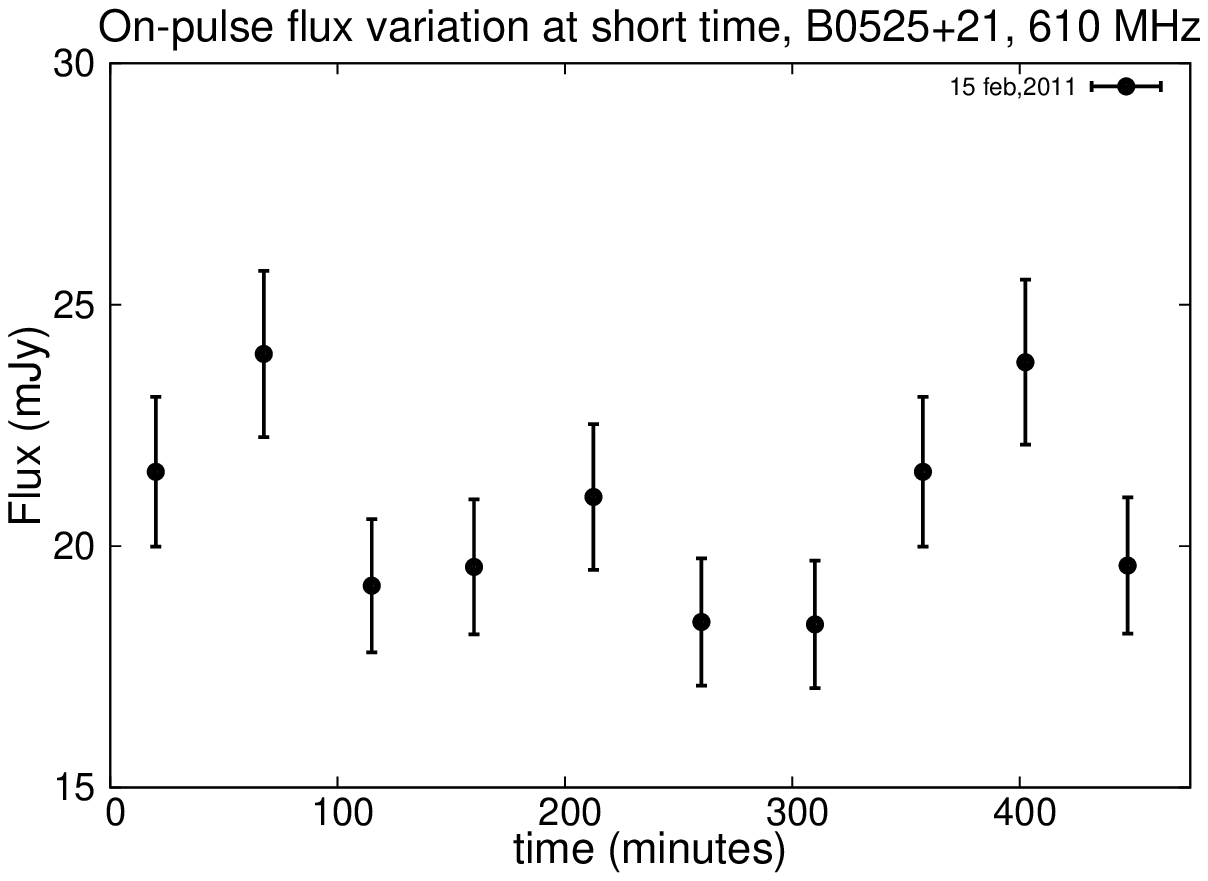}}}&
{\mbox{\includegraphics[height=4.5cm,width=7.2cm,angle=0.]{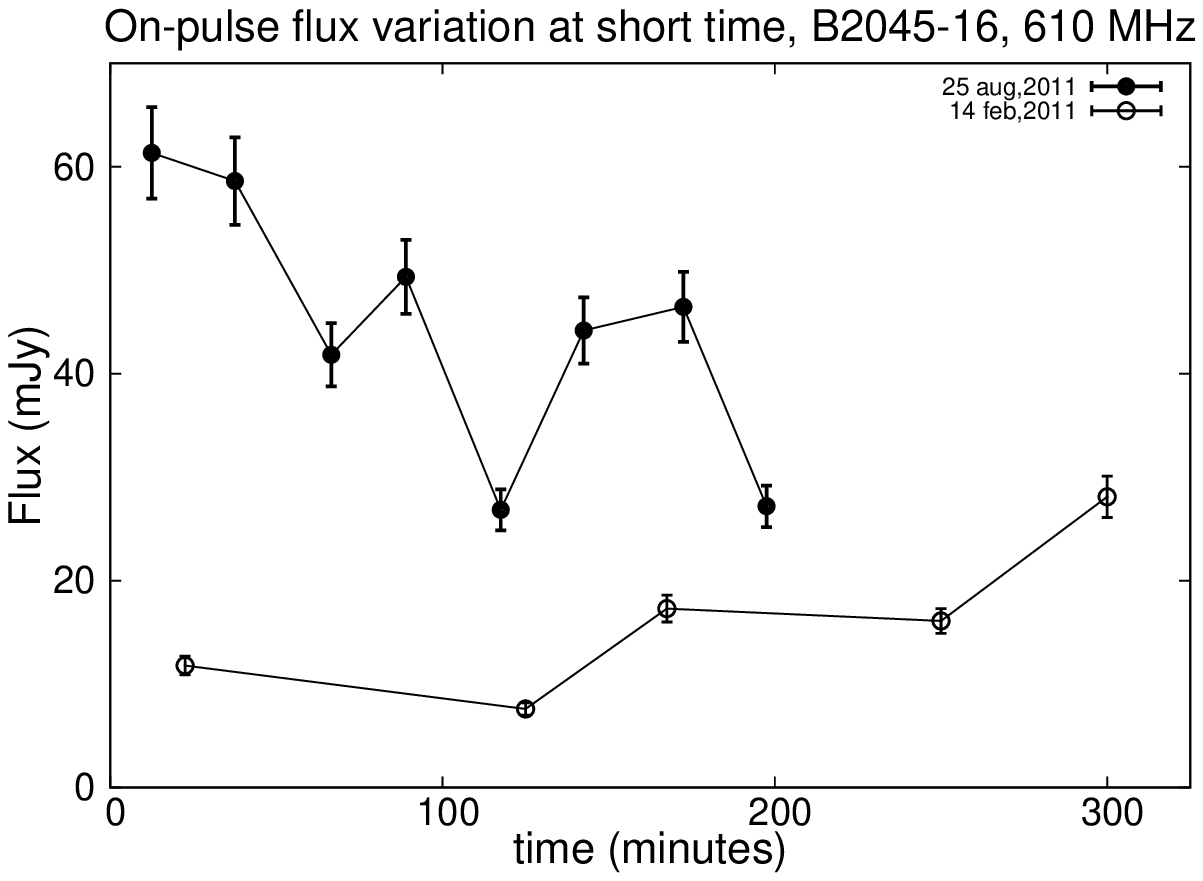}}}\\
{\mbox{\includegraphics[height=4.5cm,width=7.2cm,angle=0.]{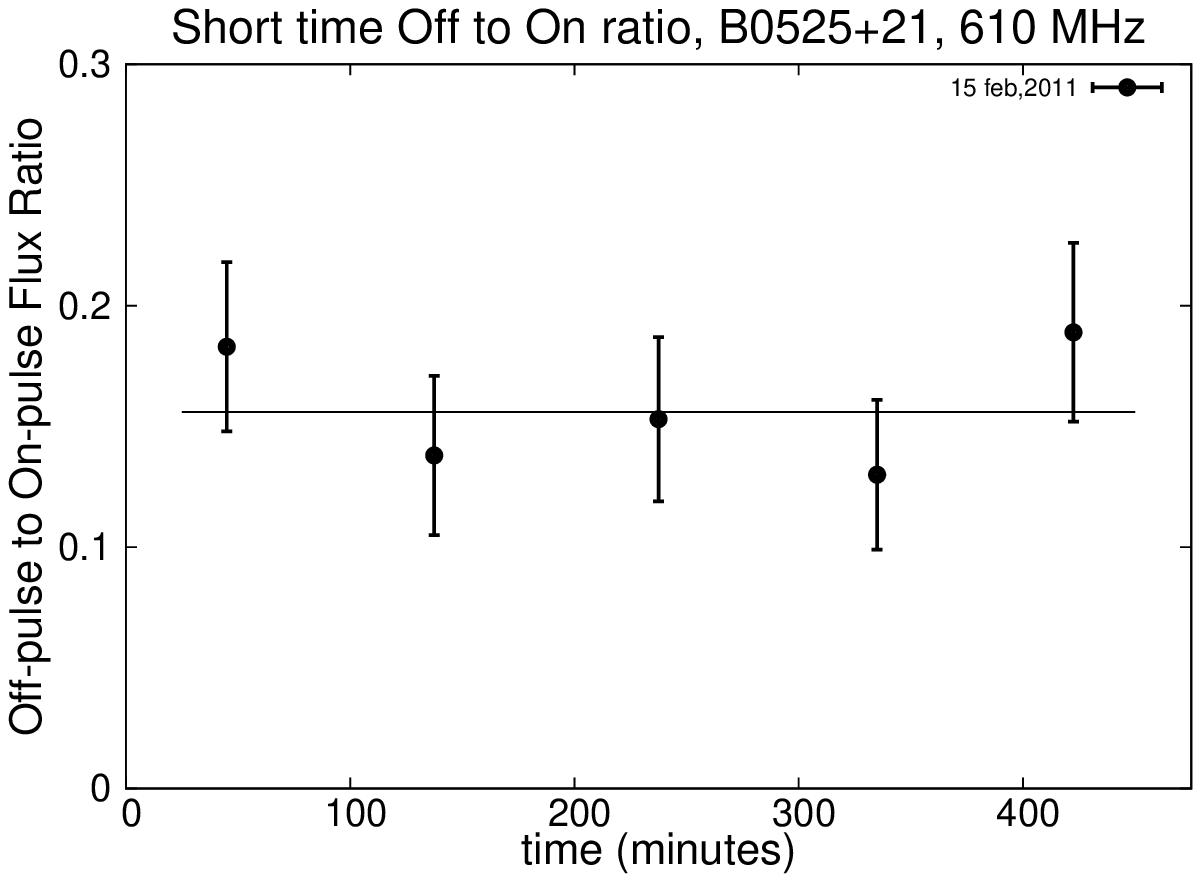}}}&
{\mbox{\includegraphics[height=4.5cm,width=7.2cm,angle=0.]{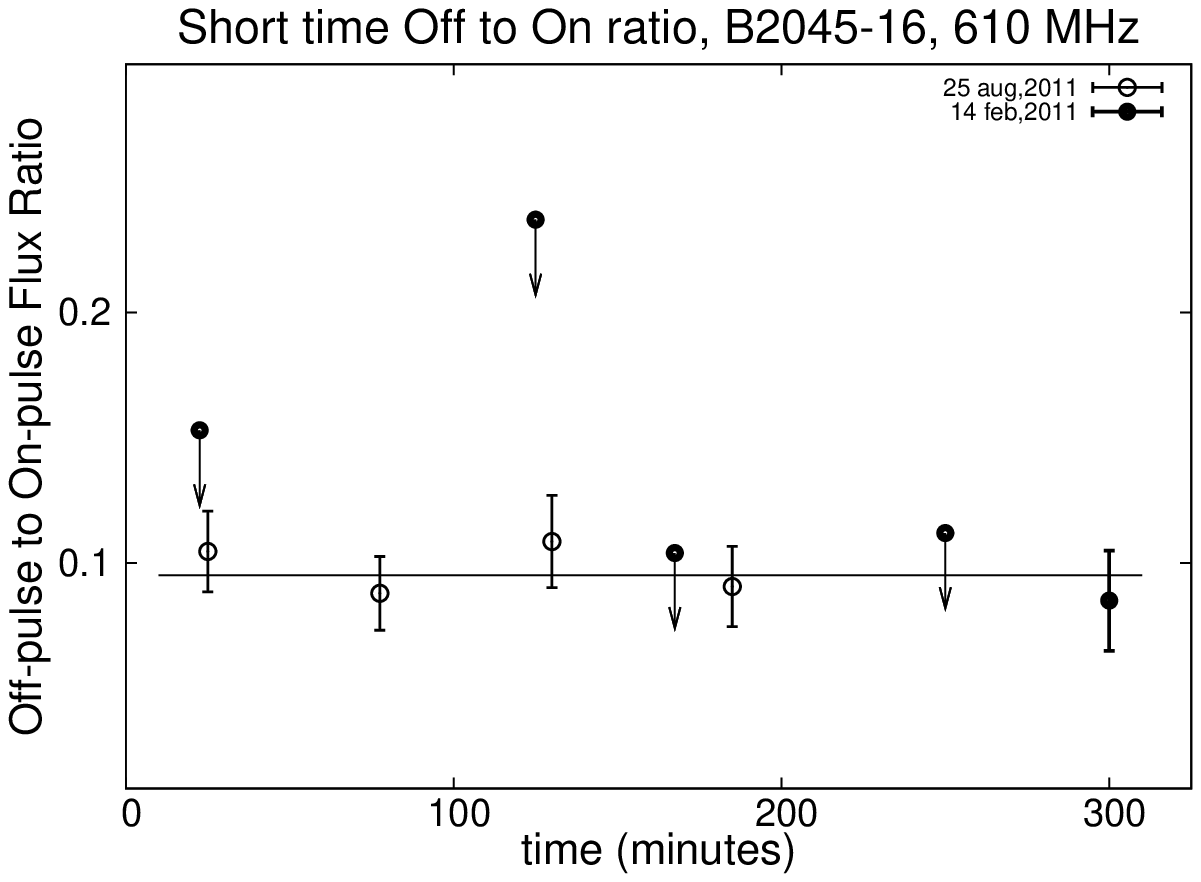}}}\\
\end{tabular}
\caption{\footnotesize The plots show the variation of the On-pulse flux and 
the Off-pulse to On-pulse flux ratio at shorter timescales. The duration of a
single observation spans several hours and the On- and Off-pulse fluxes were
determined averaging 0.5 --1 hour at a time within a single observation. 
For the pulsar B0525+21 (left) the On-pulse flux varied at short timescales
(several hours) within errors of measurements and the On-pulse to Off-pulse 
ratio remained constant at these timescales. In case of B2045--16(right) the 
On-pulse flux variations were once again within errors at 325 MHz with the 
Off-pulse to On-pulse flux ratio remaining constant. At 610 MHz the On-pulse 
flux showed large variations, however the Off-pulse to On-pulse flux ratio 
once again remained constant at short timescales for both the observations 
even though they go below detection level sometimes (arrows pointing downwards 
show upper limit of detection). This demonstrated both the long and short
timescale constancy of the Off-pulse to On-pulse flux ratio for B2045--16 at
610 MHz.}
\label{fig_short}
\end{figure*}

In paperI we argued that low energetic ($\dot{E} \sim 10^{31}$ ergs/s) long 
period ($\sim 2$ sec) pulsars B0525+21 and B2045-16, were unlikely to harbour 
PWNe around them. The ISM particle densities required to drive either a static 
or a bow shock nebula turns out to be grossly unrealistic. We concluded that 
the Off-pulse emission must have a magnetospheric origin. 

The radio emission in PWNe is due to relativistic charged particles producing 
synchrotron emission in the ambient ISM magnetic field. The typical size of 
PWNe is between 0.1--1 parsec due to expansion of the pulsar wind in the 
surrounding ISM~\citep{gae06}. If the Off-pulse emission can be constrained 
to several orders of magnitude lower than the typical PWN sizes, a PWN origin 
can be discounted. Further a wide range of studies~\citep{Weil88} suggest that 
PWNe observed in the radio frequencies between 100 MHz to 10 GHz, are 
usually characterized by a flat radio spectral index $\alpha$ in the range 
0 $> \alpha >$ -0.5 (where $S \sim \nu^{\alpha}$). The spectrum presumably 
reflects the energy spectrum of the injected particles from the pulsar. 
In some older PWNe the spectrum might steepen somewhat, like in PWNe DA495
\citep{kot08} where the spectrum steepen to $\alpha = -0.87$ above 1.3 GHz
as a result of synchrotron aging. In contrast to PWNe the pulsed emission is 
the coherent radio emission arising due to relativistically streaming charged 
particles in strong magnetic field with the observed average $\alpha \sim$ 
--1.8~\citep{mar00}. Although in some examples the spectra might show breaks 
where $\alpha$ is flatter for certain frequency range (like PSR B1952+29 has 
$\alpha = -0.6$ between 400 MHz and 1.4 GHz, also see spectra of B2045--16, 
figure~\ref{fig_spectra}), mostly they are consistent with a steep power law 
spectra and a very low frequency turnover (around 100 MHz and below). Based 
on our current understanding, any emission originating close to the pulsar 
magnetosphere and with spectral index $\alpha$ steeper than -0.9 is unlikely 
to be due to PWNe.

We set out to determine the spectral index of off-pulse emission based on 
our observations at 325 and 610 MHz. Spectral index determination requires 
robust flux estimates which is often difficult for pulsars. The measured 
pulsar flux is time variant due to both intrinsic and extraneous effects. 
The intrinsic variations occur in timescales of microseconds to several 
minutes manifesting as microstructures in single pulses, drifting subpulses, 
nulling, mode changing, etc. However the profile of a pulsar obtained 
averaging 2500--3000 pulses is relatively stable~\citep{rat95}. The pulsar 
flux is also affected by the inhomogeneities in the ISM leading to diffractive 
interstellar scintillations (DISS) and refractive interstellar scintillations 
(RISS). The DISS manifests as temporal flux variations with diffraction 
timescales ($T_{d}$) ranging from seconds to minutes and also as decorrelation 
of flux with frequency known as decorrelation bandwidth ($\Delta\nu_{d}$). 
The RISS shows variations at timescales ($T_{r}$) of days to years with a 
flux modulation index ($m_{r}$). These quantities scale with dispersion 
measure and frequency of observation typically as $T_{d} \varpropto \nu^{1.2}$,
$\Delta\nu_{d} \varpropto \nu^{4.4}$, $T_{r} \varpropto \nu^{-2.2}$ and $m_{r} 
\varpropto \nu^{0.58}$ \citep{sti90}. In order to determine the spectral 
index of the off-pulse emission it is imperative we understand its short 
and long term flux variations at each frequency.\\\\

{\bf PSR B0525+21} has a dispersion measure of 50.9 pc~cm$^{-3}$ and its flux 
has been monitored at different frequencies for several years as reported in 
Lorimer et al. (1995) and Stinebring et al. (2000) (see fig~\ref{fig_spectra}
for the average flux at low frequencies). The long term averaging results in
reduced contribution of RISS and DISS to the pulsar flux. Stinebring et al. 
(2000) quotes the scintillation parameters at our observing frequency 610 MHz 
as $T_{d} \sim$ 70 seconds, $\Delta\nu_{d} \sim$ 230 KHz, $m_{r} \sim$ 0.32, 
$T_{r} \sim$ 4 days. Using the scaling relations the equivalent quantities at 
325 MHz are $T_{d} \sim$ 33 seconds, $\Delta\nu_{d} \sim$ 15 KHz, $m_{r} \sim$ 
0.22, $T_{r} \sim$ 16 days.

Each of the GMRT observations for B0525+21 extends for about 6 to 8 hours 
and 16 MHz bandwidth at both 325 and 610 MHz. Any flux variations due to DISS
will be greatly reduced over these scales. Due to RISS one expects 5\% and 
10\% variation in flux over a single observing run at 325 MHz and 610 MHz 
respectively (these are comparable to the error in flux measurements and 
hence undetectable). In Table~\ref{Obs_detail} the average On- and Off-pulse 
flux and the Off-pulse to On-pulse flux ratio during each observing session 
is reported. The long term flux change (about 55\%) seen at 325 MHz 
(figure~\ref{fig_long}, 1$^{st}$ panel on left) is due to RISS, however the 
Off-pulse to On-pulse flux ratio remain constant both at 325 and 610 MHz 
(see figure~\ref{fig_long}, 2$^{nd}$ panel on left). The short term flux 
variations were determined by dividing a single observing session into 
shorter intervals (0.5 to 1 hour) and making images for each case. The On- 
and Off-pulse flux was measured for each interval and the Off-pulse to On-pulse
flux ratio was calculated (see figure~\ref{fig_short}, panel on left). The 
short timescale flux variations were within the errors of measurement, as 
expected, and the Off-pulse to On-pulse flux ratio was also constant.\\\\ 

{\bf B2045-16} has a dispersion measure of 11.5 pc~cm$^{-3}$ and its long 
term flux has been monitored at multiple frequencies by Stinebring \& Condon 
(1990) and Lorimer et al. (1995). Based on daily flux measurements for 43 days, 
Stinebring \& Condon (1990) quotes the scintillation parameters at 310 MHz 
(near our observing frequency of 325 MHz) as $T_{d} \sim$ 63 seconds, $\Delta
\nu_{d} \sim$ 288 KHz, $m_{r} \sim$ 0.60 and $T_{r} \sim$ 1.5 days. Using the 
scaling relations the equivalent quantities at 610 MHz are $T_{d} \sim$ 134 
seconds, $\Delta\nu_{d} \sim$ 4 MHz, $m_{r} \sim$ 0.85 and $T_{r} \sim$ 0.5 
days. 

The observations extended for 3 to 5 hours over 16 MHz bandwidth at both 325 
and 610 MHz for each observing run. Any DISS related flux variations is 
expected to reduce due to averaging over time and frequency. However due to 
RISS we expect large flux variations of 110\% and 40\% over a single 
observing run at 610 and 325 MHz respectively. In table~\ref{Obs_detail} the 
average On- and Off-pulse flux measurements and the Off-pulse to On-pulse 
flux ratios for each observation is quoted. The long term flux variations 
and Off-pulse to On-pulse flux ratio is shown in figure~\ref{fig_long}, right 
panel. The pulsar flux show variation in the flux values which are expected 
due to RISS, however the Off-pulse to On-pulse flux ratio also showed large 
variations at long timescales. The short timescale flux values were determined 
once again by dividing a single observing session into shorter intervals 
(0.5 to 1 hour) and making images for each case. The On-pulse flux at 610 MHz
showed large variations in excess of 100\% (see figure~\ref{fig_short}, 
3$^{rd}$ panel on right), however the Off-pulse to On-pulse flux ratio, when 
detected, remained at constant level for the two observations separated 
by months (see figure~\ref{fig_short}, 4$^{th}$ panel on right). The apparent 
variation of the Off-pulse to On-pulse flux ratio at large timescales can be 
attributed to the fact that the Off-pulse was below detection limit for a 
large fraction of the observing run resulting in under estimation of the 
average Off-pulse flux over the entire observing run. At 325 MHz the short 
timescale measurements once again showed constant level Of Off-pulse to 
On-pulse flux ratio at short timescales for the observation on 3 august, 2011 
(see figure~\ref{fig_short}, 2$^{nd}$ panel on right). The observation on 
16 january, 2011 was affected by the telescope pointing 34\arcmin~away from 
the pulsar resulting in increased noise levels at pulsar position thereby 
making the short interval studies impossible (the Off-pulse was undetected 
for the short interval studies due to higher noise levels). We conclude that 
despite the large variation of the On-pulse flux and the apparent variation 
of Off-pulse to On-pulse flux ratio at large timescales, the Off-pulse to 
On-pulse flux ratios remain constant at all timescales for this pulsar.

\subsection{\large Refractive Scintillations: Upper limits to emission region.}
\label{subsec3.1}

The Off-pulse to On-pulse flux ratio as demonstrated above remains constant 
at all timescales for both the pulsars. This signifies that the timescales of 
refractive scintillations for the Off-pulse is similar to that of the main 
pulse, which readily puts a constraint on the size of the emitting region of 
the Off-pulse with respect to the On-pulse. If we assume a thin screen 
approximation for refractive scintillation, i.e, the refractive scintillations 
is due to a thin lens (0.001 -- 0.005 fraction of the thickness of intervening 
medium) placed a fractional distance $\beta$ ($\beta = D_{s}/D$, where $D_{s}$ 
is separation between pulsar and lens and $D$ the distance of pulsar from 
observer) from the pulsar, the refractive timescale is given as:
\begin{equation}
 T_{r} = \frac{D_{r}}{V_{trans}\times(1 - \beta)} 
\label{eqn_Refr}
\end{equation}
where $D_{r}$ is the diameter of the lens and $V_{trans}$ the transverse 
velocity of the pulsar in the sky plane with respect to observer. The similar 
refractive timescales imply the On-pulse and Off-pulse emission is unresolved 
with respect to the lens putting an upper limit to emitting regions for the 
On- and Off-pulse with the maximum angular separation given as $\theta_{max} 
\sim \lambda / D_{r}$. 

Using the known properties of the pulsars (table~\ref{Off_size}) and assuming 
the refractive lens lies mid way between the pulsar and observer ($\beta = 
0.5$) we calculate the diameter of the lens ($D_{r}$) in each case using 
equation \ref{eqn_Refr}. This is further used to determine the upper limits 
to the angular size of the emitting region ($\theta_{max}$). Finally the 
physical size of the maximum emitting region ($R_{max}$) is calculated using 
$\theta_{max}$ and distance to the pulsars (see table~\ref{Off_size}). The 
maximum emitting region (which is also the maximum separation between the 
On-pulse and Off-pulse emitting regions) is constrained to be a few microarcsec
($\theta_{max}$). This further implies that the Off-pulse emission is 
constrained by the refractive scintillations to originate within an order 
of magnitude of the light cylinder radius ($P c /2\pi$). This as discussed 
previously is several orders of magnitude lower than the typical size of a 
PWN.

\begin{table*}
\begin{center}
\caption{The upper limits to emission region for Off-pulse emission using 
refractive scintillation.
\label{Off_size}}
\scriptsize{\begin{tabular}{ccccccccccccc}
\tableline\tableline
Pulsar & period & dist &$V_{trans}$&$T_{r}^{325}$&$T_{r}^{610}$&$P c /2\pi$&$D_{r}^{325}$&$D_{r}^{610}$&$\theta_{max}^{325}$&$\theta_{max}^{610}$&$R_{max}^{325}$&$R_{max}^{610}$\\
       &   sec  &  kpc &km~$s^{-1}$&      day      &      day      &       km         &     km        &     km        &      \arcsec         &       \arcsec        &       km        &       km        \\
\tableline
B0525+21 & 3.746& 2.28 &    229    &       16      &       4       & 1.8$\times10^{5}$&1.6$\times10^{8}$&4.0$\times10^{7}$&1.2$\times10^{-6}$&2.5$\times10^{-6}$& 4.1$\times10^{5}$& 8.6$\times10^{5}$ \\
B2045--16& 1.962& 0.95 &    511    &       1.5     &      0.5      & 0.9$\times10^{5}$&3.3$\times10^{7}$&1.1$\times10^{7}$&5.7$\times10^{-6}$&9.3$\times10^{-6}$& 8.2$\times10^{5}$& 13.2$\times10^{5}$ \\

\tableline\tableline
\end{tabular}
}
% Any table notes must follow the \end{tabular} command.
\end{center}
\end{table*}
\subsection{\large Spectral index of On- and Off-pulse emission.}
\label{subsec3.2}

\begin{figure*}
\begin{tabular}{@{}lr@{}}
{\mbox{\includegraphics[height=5cm,width=8cm,angle=0.]{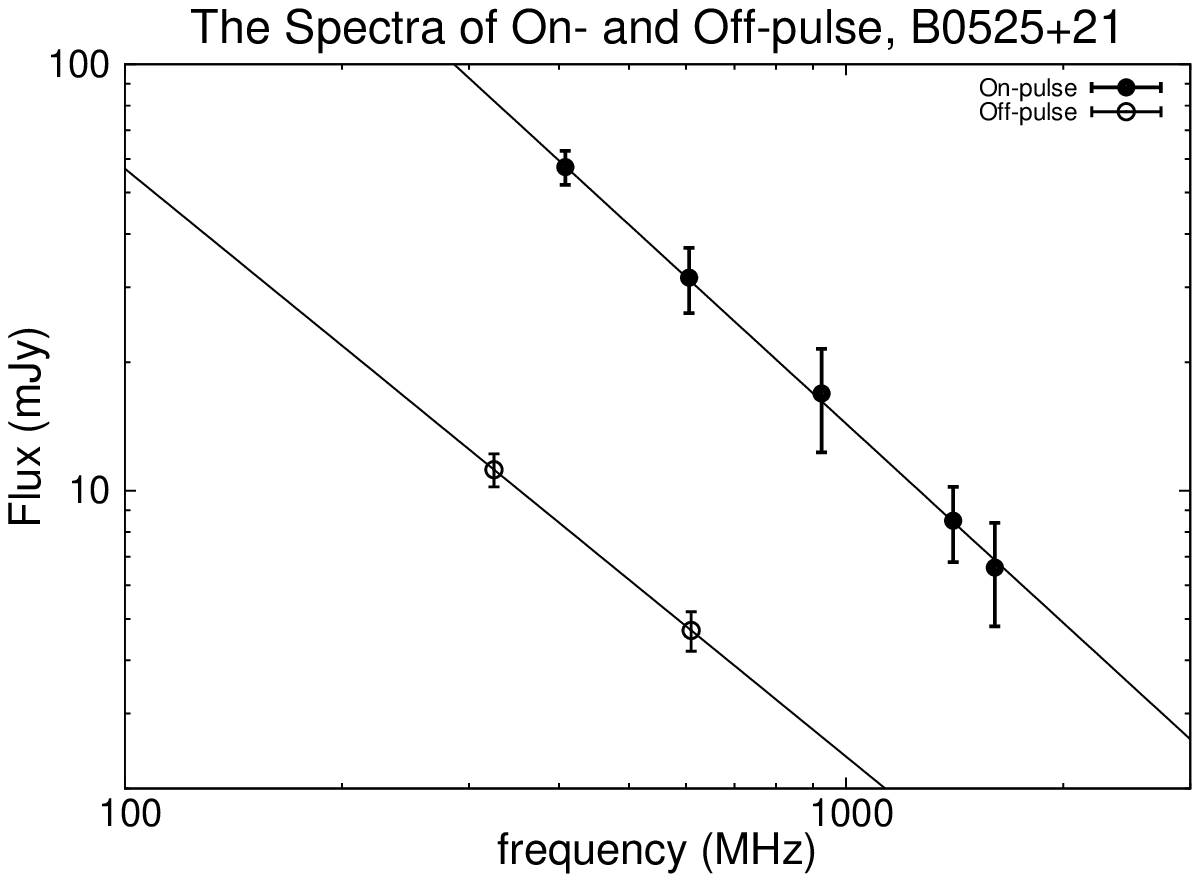}}} &
{\mbox{\includegraphics[height=5cm,width=8cm,angle=0.]{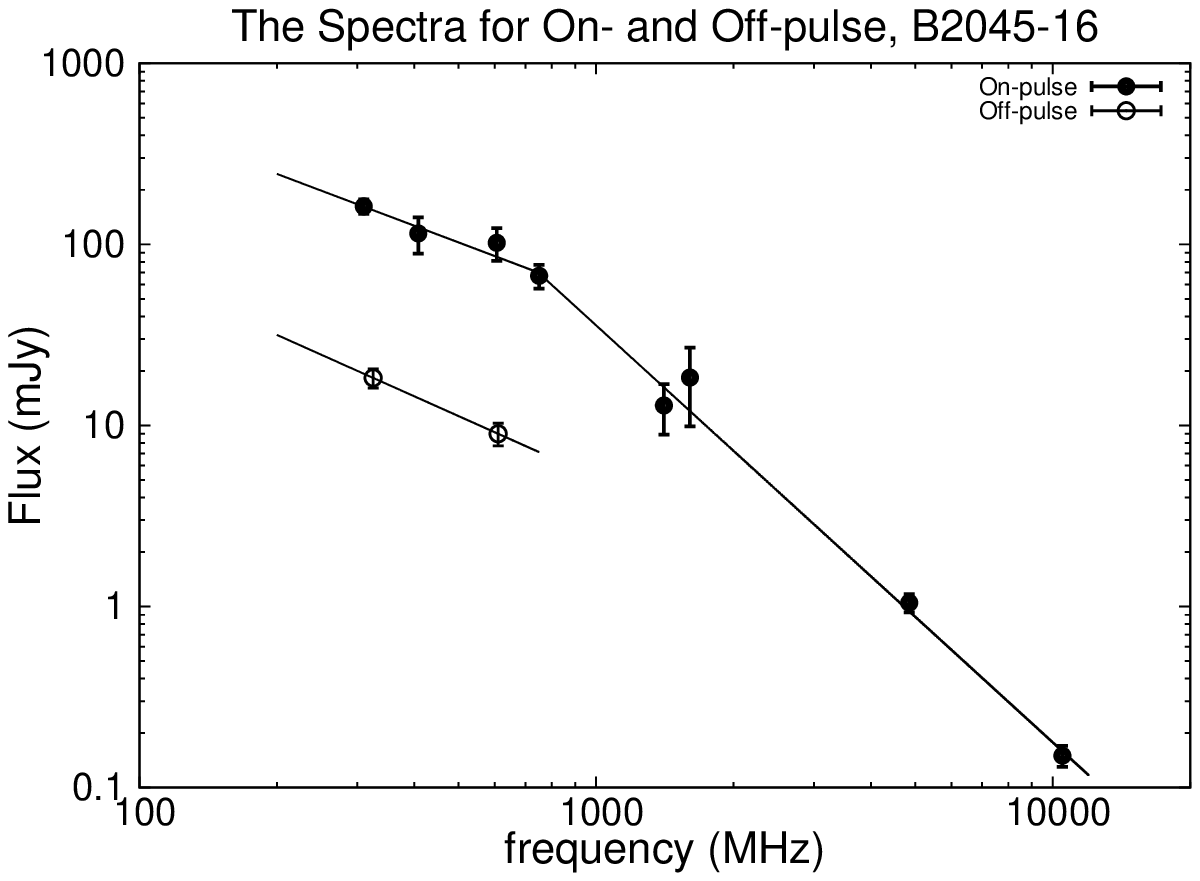}}} \\
\end{tabular}
\caption{The plots show the On-pulse spectra and the derived Off-pulse spectra 
(using equation~\ref{spect_indx}) for the two pulsars. For the pulsar B0525+21
(left) the On-pulse spectrum was determined between 300 MHz and 1.6 GHz by 
fitting a power law to the archival data~\citep{mar00, lor95}. In case of 
B2045--16(right) the On-pulse spectrum was obtained between 300 MHz and 10 
GHz. There is a break in spectrum around 750 MHz and two separate spectral 
index was calculated with values of -2.3 above 750 MHz which flattened to 
-0.95 at low frequencies.}
\label{fig_spectra}
\end{figure*}

\begin{table*}
\begin{center}
\caption{The spectral index for Off-pulse emission. The On-pulse spectra was 
determined by fitting the spectra for the flux from literature at low 
frequencies and using the Off- by On-pulse flux ratio to the Off-pulse 
spectra was calculated.
\label{spec_indx}}
\begin{tabular}{ccccc}
\tableline\tableline
Pulsar &$\alpha_{ON}$&($\frac{Off}{On})_{325}$&($\frac{Off}{On})_{610}$&$\alpha_{OFF}$ \\
       &             &                        &                        &               \\
\tableline
 B0525+21& -1.5$\pm$0.1 & 0.139$\pm$0.012 & 0.150$\pm$0.015 & -1.38$\pm$0.20\\
B2045--16&-0.95$\pm$0.12& 0.118$\pm$0.014 & 0.106$\pm$0.015 & -1.12$\pm$0.22\\
\tableline\tableline
\end{tabular}
% Any table notes must follow the \end{tabular} command.
\end{center}
\end{table*}
We have established that the Off-pulse to On-pulse flux ratio remain constant
at all timescales. This enables us to determine the Off-pulse spectral index 
between 325 MHz and 610 MHz provided the On-pulse spectral index is known and 
both the emissions follow a power law spectrum. The off-pulse spectral index 
is given as:
\begin{equation}
  \alpha_{OFF} = \alpha_{ON} + \frac{log(\rho_{1}/\rho_{2})}{log(\nu_{1}/\nu_{2})}
\label{spect_indx}
\end{equation}
where $\alpha_{OFF}$ and $\alpha_{ON}$ are the Off-pulse and On-pulse spectral
indices respectively and $\rho$ the Off-pulse to On-pulse flux ratio at 
frequency $\nu$. 

For the pulsar B0525+21 the spectrum between 400 MHz and 1.6 GHz is well known
~\citep{lor95} shown in figure~\ref{fig_spectra}, which is well approximated 
by a single power law spectrum with spectral index --1.5 (linear fit in fig
\ref{fig_spectra}). The Off-pulse spectral index is calculated using equation
\ref{spect_indx} and the Off-pulse to On-pulse flux ratios given in table
\ref{spec_indx}. The off-pulse spectral index turns out to be --1.4 which is
pretty steep and comparable to the On-pulse spectral index.
  
For the pulsar B2045--16 we once again determine the spectrum from the 
literature (Lorimer et al. 1995 and Maron et al. 2000) over the frequency
range 300 MHz to 10 GHz. In this case however the entire spectral range
is not approximated by a single power law, with a clear break in the spectrum
around 750 MHz (see figure~\ref{fig_spectra}). We determine the spectral index 
in this case by fitting two separate power laws above and below the break 
with spectral indices of --2.3 for the high frequency range and --0.95 for 
the low frequency regime. Our observations at 325 and 610 MHz lie in the low 
frequency range and we use this to calculate the spectral index of Off-pulse 
emission using equation~\ref{spect_indx} which turns out to be --1.1. This 
turns out to be steeper than the On-pulse spectra at the low frequency range.

\section{\large Discussion}
\label{section4}

The discovery of radio emission in the Off-phase of two long period pulsars 
PSR B0525+21 and B2045--16 was reported in paperI. This emission, termed 
`Off-pulse emission', occur far away from the typical main pulse in the 
pulsar profile ($\geq$ 80\degr~from the peak of main pulse). The off-pulse 
emission unlike other emissions away from the main-pulse do not appear as 
temporal structures in the pulsar profile but rather as a constant background 
emission. \footnote{A more sensitive phase resolved study in currently 
underway where we aim at imaging each phase resolved bin to localize the 
off-pulse emission region.}

In this paper we have focused on two important aspects:

(a) The off-pulse emission is unique in pulsars and it is important to 
ascertain that by no means the emission arises due to any instrumental effect 
or errors during data analysis. In section~\ref{section2} we report detection 
of the off-pulse emission for B0525+21 and B2045--16 at widely separated times 
and different frequencies. We have conducted extensive tests and ruled out 
temporal leakage from a strong pulse as it passes through the GMRT receiver 
system as a possible source of Off-pulse emission. With these tests we are 
now absolutely certain that the off-pulse emission has an astronomical origin.

(b) The next important step was to establish through observational arguments 
the nature of Off-pulse emission. In section~\ref{section3} we demonstrate 
that for PSR B0525+21 the Off-pulse to On-pulse flux ratio is constant at 
both 325 and 610 MHz, which in turn gives a spectral index $\alpha_{OFF} 
\sim$ --1.4. We have also demonstrated that despite apparent variations in 
the long term Off-pulse to On-pulse flux ratio for the pulsar B2045--16, it 
actually remains constant at all timescales when the Off-pulse flux variations 
are taken into account. This once again allowed us to determine the spectral 
index of $\alpha_{OFF} \sim$ --1.1. Based on the observed properties of 
refractive scintillation, we derived the radio emission region of the Off-pulse 
emission to have a maximum size of magnetospheric scale (see 
section~\ref{subsec3.1}). The steep spectral index coupled with a highly 
compact emission region make it highly unlikely for the Off-pulse emission to
be a PWNe.

The Off-pulse emission appear to be a completely new and hitherto undetected 
magnetospheric emission from pulsars. The estimated brightness temperature 
of the Off-pulse emission is greater than 10$^{18}$ K, assuming the emission 
to originate at the light cylinder. This strongly suggest a coherent radio 
emission process as the mechanism for explaining the Off-pulse emission which 
leads to the next important question as to where and how does this off-pulse 
coherent emission originate in the pulsar magnetosphere? Currently we do 
not have any good answers to these questions but there are a few likely 
possibilities that can be considered. The basic pulsar models suggest that 
the power from the rotational energy in a neutron star is tapped and converted 
into the observed radiation. The rotating magnetic field of the neutron star 
acts like a unipolar inductor, creating high electric fields around the
neutron star where charged particles are accelerated to relativistic 
velocities along the open field lines. This eventually leads to the pulsar 
radiation. A large number of models exist which try to establish the location 
of the charge accelerating regions and suggest physical mechanisms that can 
excite the coherent radio emission in pulsars. However here we will not go 
into the details of any of these models, but would like to point out as
why finding the location of the emission region in the pulsar magnetosphere 
is central in unravelling the origin of the Off-pulse emission.

In most pulsar theories the radio emission is excited due to development 
of plasma instabilities in the outflowing plasma. The main pulse emission 
arises around 500 km above the neutron star surface \citep{bla91, ran93, 
kij98, krz09}. At these heights the magnetic field is very strong, and the 
relativistic charged plasma particles are forced to move along the curved 
magnetic field lines. Here the well known two-stream plasma instability 
naturally generates langmuir plasma waves. If the plasma is subjected to a 
non-stationary flow, then it leads to the modulational instability of Langmuir 
waves which results in formation of relativistic charged solitons capable of 
emitting coherent curvature radiation~\citep{mel00, gil04, mit09}. The Off-
pulse emission on the other hand lies in the region outside the main pulse, 
and currently we are not certain about its exact location. For the emission 
to originate from open field lines, the location needs to be above the main 
pulse in regions where the open dipolar magnetic field lines diverge. As the 
pulsar rotates, this divergent region above the main pulse is sampled by the 
observer as the Off-pulse emission. The other well known maser type plasma 
instability that can give rise to the coherent emission are the cyclotron--
cherenkov or cherenkov drift instability~\citep{kaz87, kaz91}. The 
instabilities are due to the interaction of the fast particles of the primary 
beam with the normal modes of the electron-positron plasma. These instabilities
are known to operate close to the light cylinder and could be a viable source 
of Off-pulse emission.

There might be other possible origin of the Off-pulse emission which we will 
need to evaluate and understand in the future, however it is clear that this 
newly found Off-pulse emission from low energetic slowly rotating neutron 
stars provide important clues in understanding the physical phenomenon 
operating in the pulsar magnetosphere.

\section{\large Acknowledgments}
We would like to thank Jayanta Roy for developing the 128 millisecond 
interferometric mode of the GMRT software backend(GSB) which were used 
as a part of these observations. We would also like to thank Navnath 
Shinde, Sweta Gupta, Ajay Vishwakarm, Ajit Kumar and other members of 
the GMRT engineering team for developing the pulsed noise source used in
our studies. We thank the staff of the GMRT that made these observations 
possible. GMRT is run by the National Centre for Radio Astrophysics of 
the Tata Institute of Fundamental Research.

{\it Facility:} \facility{GMRT}.

\clearpage


\begin{thebibliography}{}

\bibitem[Basu et al. 2011]{bas11} Basu, R., Athreya, R., Mitra D.  2011,
    \apj, 728, 157
\bibitem[Blaskiewicz et al. 1991]{bla91} Blaskiewicz, M.; Cordes, J.M.; 
    Wasserman, I.  1991, \apj, 370, 643
\bibitem[Frail \& Scharringhausen 1997]{fra97} Frail, D.A., 
    Scharringhausen, B.R.  1997, \apj, 480, 364
\bibitem[Gaensler \& Slane 2006]{gae06} Gaensler, B.M.; Slane, P.O.  2006,
    \araa, 44, 17
\bibitem[Goldreich \& Julian 1969]{gol69} Goldreich, P., Julian, W. H.  1969,
    \apj, 157, 869
\bibitem[Gil et al. 2004]{gil04} Gil, J.; Lyubarsky, Y.; Melikidze, G.I.  2004,
    \apj, 600, 872
\bibitem[Kazbegi et al. 1987]{kaz87} Kazbegi, A.Z.; Machabeli, G.Z.; 
    Melikidze, G.I.  1987, AuJPh, 40, 755
\bibitem[Kazbegi et al. 1991]{kaz91} Kazbegi, A.Z.; Machabeli, G.Z.; 
    Melikidze, G.I.  1991, \mnras, 253, 377
\bibitem[Kijak \& Gil 1998]{kij98} Kijak, J.; Gil, J.  1998, \mnras, 299, 855
\bibitem[Kothes et al. 2008]{kot08} Kothes, R.; Landecker, T.L.; Reich, W.;
    Safi-Harb, S.; Arzoumanian, Z.  2008, \apj, 687, 516
\bibitem[Krzeszowski et al. 2009]{krz09} Krzeszowski, K.; Mitra, D.; Gupta, Y.;
    Kijak, J.; Gil, J.; Acharyya, A.  2009, \mnras, 393, 1617
\bibitem[Lorimer et al. 1995]{lor95} Lorimer, D.R., Yates, J.A., Lyne, A.G.,
    Gould, D.M.  1995, \mnras, 273, 411
\bibitem[Maron et al. 2000]{mar00} Maron, O.; Kijak, J.; Kramer, M.;
    Wielebinski, R.  2000, \aaps, 147, 195
\bibitem[Melikidze et al. 2000]{mel00} Melikidze, G.I.; Gil, J.A.; 
    Pataraya, A.D.  2000, \apj, 544, 1081
\bibitem[Mitra et al. 2009]{mit09} Mitra, D.; Gil, J.; Melikidze, G.I.  2009, 
    \apj, 696, 141
\bibitem[Rankin 1993]{ran93} Rankin, J.M.  1993, \apj, 405, 285
\bibitem[Rathnasree \& Rankin 1995]{rat95} Rathnasree, N., Rankin, J. M.  1995,
    \apj, 452, 814
\bibitem[Rickett 1990]{ric90} Rickett, B.J.  1990, \araa, 28, 561
\bibitem[Ruderman \& Sutherland 1975]{rud75} Ruderman, M. A.,
    Sutherland, P. G.  1975, \apj, 196, 51
\bibitem[Stinebring \& Condon 1990]{sti90} Stinebring, D.R., Condon, J.J,
    1990, \apj, 352, 207
\bibitem[Stinebring et al. 2000]{sti00} Stinebring, D.R., Smirnova T.V.,
    Hankins, T.H., Hovis, J.S., Kaspi, V.M., Kempner, J.C., Myers E.,
    Nice D.J.  2000, \apj, 539, 300
\bibitem[Thompson et al. 1986]{tho86} Thomson, A.R., Moran, J.M.,
    Swenson G.W.  1986, Interferometry and Synthesis in Radio Astronomy
    (Wiley Interscience, New York).
\bibitem[Weiler \& Sramek 1988]{Weil88} Weiler, K.W.; Sramek, R.A.,
    1988, \araa, 26, 295

\end{thebibliography}
\end{document}